\documentclass[12pt]{article}
\usepackage[dvips]{graphicx}
\usepackage{pdproc}

  \makeatletter 
  \def\@cite#1{[#1]} 
  \makeatother    
\begin{document}

\renewcommand{\thefootnote}{\alph{footnote}}

\title{
Fermion Masses and Bi-large Lepton Mixing from Singular Matrices
}

\author{ RADOVAN DERM\'I\v SEK}

\address{ 
Davis Institute for High Energy Physics,  
University of California \\
Davis, California 95616, USA
\\ {\rm E-mail: dermisek@physics.ucdavis.edu}}

\abstract{
We discuss conditions under which bi-large mixing
in the lepton sector is achieved 
assuming that
all Yukawa matrices and
the right-handed
neutrino  mass matrix have the same singular form in the leading order.
Due to obvious quark lepton symmetry, this approach can be embedded into grand 
unified theories.
The right-handed neutrino mass scale can
be identified with the GUT scale and, as a consequence of third generation 
Yukawa coupling unification, 
the mass of the lightest neutrino
is given as $(m_{top}^2/M_{GUT}) \, | \, U_{\tau 1} \, |^2$ in the leading order.
This relation  does not depend on 
the exact form of mass matrices.
}

\normalsize\baselineskip=15pt

\section{Introduction}

In order to keep quark-lepton symmetry obvious, let us assume that all mass matrices are given
by the same universal matrix in the leading approximation, and all differences between up-type 
quarks, down-type quarks, charged leptons and neutrinos originate from small perturbations.
Examples of such universal matrices are democratic mass matrix, a matrix with 3-3 element 
only, or, in general, a matrix formed by a product of two vectors:
\begin{equation} 
{\cal I}_D =       \left(\begin{array}{ccc}  1 & 1 & 1 \\
                                           1 & 1 & 1 \\
                                           1 & 1 & 1
                 \end{array} \right), 
\quad \quad
{\cal I}_{33} =       \left(\begin{array}{ccc}  0 & 0 & 0 \\
                                           0 & 0 & 0 \\
                                           0 & 0 & 1
                 \end{array} \right),
\quad \quad
{\cal I}_{LR} \, = \, {\vec \mu}_L \, . \, {\vec \mu}^{\, T}_R .
\end{equation}
The democratic mass matrix ${\cal I}_D$ does not distinguish between families, all three families 
are treated equally in the leading order, and yet it provides an explanation for the third 
generation being of order the weak scale and the first two generations being massless.
If embedded into grand unified theories third generation Yukawa coupling unification is a generic 
feature (without necessity of distinguishing the third generation from the other two by
family
symmetries or in any other way) while masses of the first two generations of charged fermions depend 
on
small perturbations. A matrix with non-zero 3-3 element only distinguishes the third 
generation from 
the beginning and it is the usual starting point of hierarchical models. I will mainly focus on the 
first example following Ref.~\cite{dem} (for a related study, see Ref.~\cite{branco_efd}) and I will 
comment on similarities and
differences in corresponding hierarchical models~\cite{3gd}. 
However, the discussion and many of the consequences are similar for any matrix which can be 
written as a product of two vectors, ${\cal I}_{LR}$. (For a specific example with $ {\vec \mu}_L = 
{\vec \mu}_R = (\lambda^2, \lambda, 1 )^T$ see Ref.~\cite{dorsner_smirnov}). 
Such mass matrices can originate from an exchange of a heavy vector-like pair of fermions
where ${\vec \mu}_L$ (${\vec \mu}_R$) are couplings of left-handed (right-handed) fermions. 
The former two examples are clearly just special cases. 
Especially in the case $ {\vec \mu}_L \sim
{\vec \mu}_R \sim (1, 1 , 1 )^T$, in other words when all elements are random order one numbers, the 
results that follow are basically identical to the case of ${\cal I}_D$. 

\newpage

\section{Bi-large lepton mixing in a democratic approach}

Let us assume that Yukawa couplings are given as:
\begin{equation}
Y_f \equiv \frac{1}{3} \lambda_f \left( {\cal I}_D - {\cal E}_f \right) ,  \quad \quad f = u, d, e, \nu ,
\end{equation}
where we parametrize departures from
universality by matrices ${\cal E}_f$.
If  Yukawa matrices were equal to ${\cal I} \lambda_f /3 $ the mass 
eigenvalues are
$\{ 0,0,\lambda_f \}$ and the diagonalization matrix is:
\begin{equation}
U_{\cal I} =
\left(\begin{array}{ccc}
        \ \  \cos \theta_{\cal I}   & \sin \theta_{\cal I}     & \ \ 0 \\
        - \sin \theta_{\cal I} & \cos \theta_{\cal I}     & \ \ 0 \\
       \ \   0               & 0                 & \ \ 1
   \end{array} \right) \,
            \left(\begin{array}{rrc}
      \frac{1}{\sqrt{2}} & -\frac{1}{\sqrt{2}} &  \ \ \ 0 \\
      \frac{1}{\sqrt{6}} &  \frac{1}{\sqrt{6}} & -\frac{2}{\sqrt{6}}\\
      \frac{1}{\sqrt{3}} &  \frac{1}{\sqrt{3}} &  \ \ \frac{1}{\sqrt{3}}
                 \end{array} \right)  .
\label{eq:UI3}
\end{equation}
As a consequence of  degenerate zero eigenvalues the first two rows of 
this matrix are not uniquely
specified and are model dependent ($\cal E$ has to be taken into account). 
They can be replaced by
any of their linear combinations and the corresponding orthogonal 
combination, which is accounted for by
the first matrix which rotates the first two rows.
As a result, the CKM matrix is not the identity
matrix in the leading order as it was in the case of two families, but 
rather a unitary matrix
with an arbitrary
1-2 element.

Let us parametrize the Majorana mass matrix for right-handed neutrinos in a 
similar way:
\begin{equation}
M_{\nu_R} = \frac{1}{3} \left( {\cal I}_D - {\cal R} \right) M_0 , 
\end{equation}
where {\cal R} represents small perturbations.
The inverse of this matrix is given as:
\begin{equation} 
M_{\nu_R}^{-1} = \frac{1}{M_{eff}} \left( \hat{\cal I} + \hat{\cal R} 
\right),
\end{equation}
where $M_{eff} \simeq r M_0 /3$ with $r \equiv \sum_{i,j = 1}^3 \hat{\cal R}_{ij}$.
The form of $\hat{\cal I}$ can found in Ref.~\cite{dem} and $\hat{\cal R}$ contains higher 
order terms.

Due to the special form of ${\cal I}$ and $\hat{\cal I}$ we have these relations: ${\cal I} \hat{\cal 
I} = 0$, ${\cal I} \hat{\cal R} {\cal I} = r$ and the usual see-saw formula for the left-handed neutrino 
mass matrix, $M_{\nu_L} = - v^2_\nu Y_\nu M_{\nu_R}^{-1} Y_\nu^T $, highly simplifies:
\begin{equation}
M_{\nu_L} = - \frac{\lambda^2_\nu v^2_\nu}{9 M_{eff}}
 \left[ {\cal M} + r {\cal I} + O(\hat{\cal R}_{ij} \epsilon_{\nu ij} )
 \right] ,
\label{eq:MnuL3}
\end{equation}
where ${\cal M} \equiv {\cal E}_\nu \hat{\cal I} {\cal E}_\nu^T $ and we assume 
${\cal R}_{ij}$ are 
much smaller than ${\cal E}_{\nu ij}$ (so the terms 
$O(\hat{\cal R}_{ij} \epsilon_{\nu ij} )$ are negligible).
If the second term in Eq.~(\ref{eq:MnuL3}) dominates, the neutrino mass matrix resembles the charged 
lepton mass matrix and the lepton mixing matrix would be similar to the CKM matrix. In order to get 
large mixing in the lepton sector this term simply cannot dominate. On the other hand, the first term 
in Eq.~(\ref{eq:MnuL3}), ${\cal M}$, is given in terms of perturbations only. If this term dominates 
(this situation require strong hierarchy in masses of right handed neutrinos and negligible contribution of
the heaviest one to the left-handed neutrino mass matrix, $M_1 < M_2 < 10^{-4} M_3$)
there is 
absolutely no reason why the neutrino mass matrix should resemble the charge lepton mass matrix. It 
can be anything. 
In general, matrix ${\cal M}$ has one zero eigenvalue and the corresponding eigenvector, ${\vec v}_0$, 
is specified 
by ${\cal E}_\nu$ only. 

The matrix diagonalizing the charged fermion mass matrix is given (up to small corrections) by 
$U_{\cal I}$ in 
Eq.~(\ref{eq:UI3}).
Since it already 
contains large mixing angles, 
in order to
avoid any exact relations between elements of ${\cal E}_e$, ${\cal E}_\nu$ and ${\cal R}$,
the simplest
way is to assume that the perturbation matrix ${\cal E}_\nu$ introduces the minimal amount of mixing
into the lepton mixing matrix. This corresponds to a situation when the eigenvector corresponding to 
the zero eigenvalue is dominated by one element.
The most general form of the lepton mixing matrix in this
case can be written as: 
\begin{equation} U = \left(\begin{array}{ccc}
        \cos \theta_e   & \sin \theta_e     & 0 \\
        - \sin \theta_e & \cos \theta_e     & 0 \\
        0               & 0                 & 1
   \end{array} \right)
\left(\begin{array}{rrc}
      \frac{1}{\sqrt{2}} & -\frac{1}{\sqrt{2}} &  \ \ \ 0 \\
      \frac{1}{\sqrt{6}} &  \frac{1}{\sqrt{6}} & -\frac{2}{\sqrt{6}}\\
      \frac{1}{\sqrt{3}} &  \frac{1}{\sqrt{3}} &  \ \ \frac{1}{\sqrt{3}}
                 \end{array} \right)
   \left(\begin{array}{ccc}   
        1 & 0   & 0 \\
        0 & \cos \theta_\nu     & \sin \theta_\nu \\
        0 & - \sin \theta_\nu   & \cos \theta_\nu
   \end{array} \right).
\label{eq:U3}
\end{equation}
where $\cos \theta_e$ and $\cos \theta_\nu$ are free parameters. 
Plots of their allowed values  that satisfy $3 \sigma$ experimental bounds of $\sin^2 
\theta_{23}$ and $\sin^2 \theta_{12}$ can be found in~Ref.~\cite{dem}. Since the lepton mixing matrix 
is determined by only two parameters in this minimal case the value of the remaining mixing angle 
is a prediction. 
The predicted values of $\sin^2 \theta_{13}$ are either $0.008 \leq  \sin^2 \theta_{13} 
\leq 0.14$ or $0.22 \leq \sin^2 \theta_{13} \leq 0.66$. Therefore this framework naturally leads to 
either all three mixing angles large or at most one small mixing angle.
Note that the minimal value of $\sin^2
\theta_{13} = 0.008$ corresponds to the maximal allowed values of $\sin^2 \theta_{23}$ and
$\sin^2 \theta_{12}$. On the other hand, the central values of $\sin^2 \theta_{23}$ and
$\sin^2 \theta_{12}$ correspond to $\sin^2 \theta_{13}$ near its present experimental upper bound.

\section{Mass of the lightest neutrino}

The mass of the lightest neutrino is lifted when the second term in Eq.~(\ref{eq:MnuL3}) is taken into
account.
Since we assume it is just a small correction to the first two terms it can be treated as a
perturbation. Adding this perturbation does not significantly affect the two heavy eigenvalues and
the diagonalization matrix, but it is crucial for the lightest eigenvalue which is exactly zero in
the limit when this term is ignored.
In the case of non-degenerate eigenvalues, corrections to eigenvalues $m_i$ of a matrix $M$ generated
by a matrix $\delta M$ are given as $\delta m_i = u_i^\dag \, \delta M \, u_i$, 
where $u_i$ are normalized eigenvectors.
Due to the universal form of the perturbation matrix
we have ${\vec v}_0^{\, \dag} \, r {\cal I} \, {\vec v}_0 = r \, | \, \xi^2 \, |$, where $\xi = \sum_{i=1}^3 
v_{0i}$.
Therefore, the mass of the lightest neutrino is given as:   
\begin{equation}
m_{\nu_1} =  \frac{\lambda^2_\nu v^2_u}{3M_0} \, | \, \xi \, |^2 .
\label{eq:m_nu1}
\end{equation}  
The parameter $\xi$ is however related to the 3-1 element of the lepton mixing matrix, 
$U_{\tau 1} = \left( U_e U_{\nu_L}^\dag \right)_{31} = (1,1,1) . {\vec v}_0 \, / \sqrt{3} = \xi / \sqrt{3}$,
and so we get
\begin{equation}
m_{\nu_1} =  \frac{\lambda^2_\nu v^2_u}{M_0} \, | \, U_{\tau 1} \, |^2 .
\label{eq:m_nu1a}
\end{equation}  
Note, the 3rd row in $U_e$ {\it is not} model dependent unlike the first two rows are! It can receive only small
corrections from the perturbation matrix.


\newpage

In
simple SO(10)  models $\lambda_u  = \lambda_\nu $, in which case the lightest and the
heaviest fermion of the standard model
are connected through the relation above where $\lambda_\nu^2 v_u^2$ is replaced by
$m_{top}^2$. 
This is a very pleasant feature since
we can further
identify $M_0$ with the GUT scale, $M_{GUT} \sim 2 \times 10^{16}$ GeV, in which case we get
\begin{equation}
m_{\nu_1} = \frac{m_{top}^2}{M_{GUT}} \, | \, U_{\tau 1} \, |^2 ,
\label{eq:m_nu1c}
\end{equation}   
and predict the mass of the lightest neutrino
to be between $ 5 \times 10^{-5}$ eV and  $ 5 \times 10^{-4}$ eV depending on the value of $U_{\tau 1}$ 
(a global analysis of neutrino oscillation data~\cite{Gonzalez-Garcia:2003}
gives the $3 \sigma$ range: $0.20 \leq  | \, U_{\tau 1} \, |  \leq 0.58$).
This prediction does not depend on details of a model and is well motivated possible consequence of Yukawa coupling 
unification. It adds to predictions of Yukawa coupling unification in quark and charged lepton 
sector~\cite{yukawa}.

\section{Conclusions}

The scenario we discussed,
when embedded into GUTs,
is very compact and has many virtues: obvious quark-lepton 
symmetry, 3rd generation Yukawa coupling
unification, bi-large lepton mixing
with a prediction for $\sin \theta_{13}$ in the minimal case,
and more importantly,
no need for an intermediate right-handed neutrino scale and with that associated prediction for the mass of the
lightest neutrino. 

Many features of this scenario are similar to those in the hierarchical framework discussed in Ref.~\cite{3gd}.
The third generation Yukawa coupling unification is obvious in that picture.
This can be understood from two possible ways  permutation symmetry can 
be used in
model building. A matrix with 3-3 element only can be also motivated by permutation symmetry  under
which the first two families transform as a doublet~\cite{D3}.
Both approaches require strong hierarchy in masses of right handed neutrinos and negligible contribution of the 
heaviest one to
the left-handed neutrino mass matrix.


\section{Acknowledgments}

This work was supported, in part, by the U.S.\ Department of Energy, Contract
DE-FG03-91ER-40674 and the Davis Institute for High Energy Physics.

\bibliographystyle{plain}

\begin{thebibliography}{99}
\vspace{-0.2cm}

\bibitem{dem}
R.~Dermisek,
{\it Phys.\ Rev.} {\bf D70}, 033007 (2004).

\bibitem{branco_efd}
E. Kh. Akhmedov, G.C. Branco, F.R. Joaquim and J.I. Silva-Marcos, {\it Phys.\ Let.} {\bf B498}, 237
(2001).

\bibitem{3gd}
R.~Dermisek,
arXiv:hep-ph/0406017.

\bibitem{dorsner_smirnov}
I.~Dorsner and A.~Y.~Smirnov,
arXiv:hep-ph/0403305.

\bibitem{Gonzalez-Garcia:2003}
M.~C.~Gonzalez-Garcia and C.~Pena-Garay,
{\it Phys.\ Rev.} {\bf D68}, 093003 (2003).


\bibitem{yukawa}
T.~Blazek, R.~Dermisek and S.~Raby, {\it Phys.\ Rev.\ Lett.} {\bf 88}, 111804 (2002);
%
{\it Phys.\ Rev.} {\bf D65}, 115004 (2002), and references therein.

\bibitem{D3}
R.~Dermisek and S.~Raby,
{\it Phys.\ Rev.} {\bf D62}, 015007 (2000).
%
\end{thebibliography}

\end{document}